\documentclass[a4paper,onecolumn]{article}
\usepackage{graphicx}
\title{Narrow bandwidth interference filter-stabilized diode laser systems for the
manipulation of neutral atoms}
\author{M. Gilowski, Ch. Schubert, M. Zaiser, W. Herr, T. W{\"u}bbena,\\T. Wendrich, T. M{\"u}ller, E.~M.~Rasel, and W.~Ertmer}
\date{\emph{Institut f{\"u}r Quantenoptik, Welfengarten 1, 30167 Hannover,
Germany}}
\begin{document}
   \maketitle
\begin{abstract}
We present and investigate different external cavity diode laser
(ECDL) configurations for the manipulation of neutral atoms,
wavelength-stabilized by a narrow-band high transmission
interference filter. A novel diode laser, providing high output
power of more than 1~W, with a linewidth of less than 85~kHz,
based on a self-seeded tapered amplifier chip has been developed.
Additionally, we compare the optical and spectral properties of
two laser systems based on common laser diodes, differing in their
coating, as well as one, based on a distributed-feedback (DFB)
diode. The linear cavity setup in all these systems combines a
robust and compact design with a high wavelength tunability and an
improved stability of the optical feedback compared to diode laser
setups using diffraction gratings for wavelength discrimination.
\end{abstract}
\section{Introduction}
\label{intro} Diode laser systems have become an attractive light
source with versatile applications in many fields of modern physics,
such as telecommunication or the manipulation of atoms. The atom
optical experiments in the field of e.g. quantum degenerated
gases~\cite{Ketterle} or metrology~\cite{Berman} with their future
space-based experiments~\cite{Salomon,Vogel,Jentsch} make high
demands on the laser systems. The challenge is to design compact and
robust laser configurations offering a narrow linewidth and a high
output power.

In this article, we compare four different laser systems, based on
a narrow-band high transmission interference filter, as presented
in~\cite{Zorabedian} and~\cite{Baillard}. The laser designs offer
an improved stability and tunability compared to grating-based
setups combined with convincing spectral properties. In
particular, we study a novel laser design, based on a self-seeded
tapered amplifier, in the following called tapered laser (TL). The
TL provides a high output power and a narrow spectral bandwidth
combined with a higher stability, yielding a better performance
than self-seeded tapered amplifier lasers using a grating for
wavelength discrimination. In addition, the TL offers a simplified
setup compared to the well-established
master-oscillator-power-amplifier (MOPA)-system~\cite{Voigt}.
Furthermore, we study three ECDL-systems which differ in their
implemented medium power laser diode ($<$100mW), thus leading to
different properties, such as wavelength tunability and output
power of the lasers. We implemented two common laser diodes, with
and without an AR-coating, as well as a DFB-diode in the
ECDL-systems.

The article is organized as follows: We begin in
section~\ref{sec:setup} with the general description of the
studied laser configurations. The characterization of the
ECDL-setups follows in section~\ref{sec:3Laser}. Finally, we
discuss the properties of the new high output power laser
prototype in section~\ref{sec:TLLaser}.
\section{Laser
Configurations}\label{sec:setup} 
\begin{figure}[t]
\centering {
  \includegraphics[width=11cm]{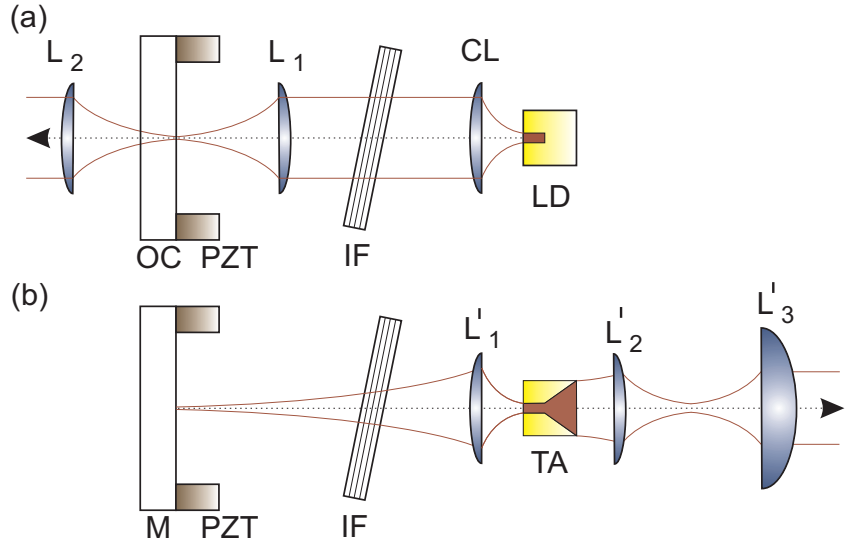}
} \caption{Schematic drawing of the laser configurations 1-3 in
(a) and of the tapered laser in (b). The optical elements laser
diode (LD) or tapered amplifier (TA), collimation lens (CL),
interference filter (IF), out-coupler (OC) or mirror (M),
piezo-electric transducer (PZT) and lenses (L$_{1,2}$) as well as
(L'$_{1,2,3}$) for the cat eye configuration and collimation
purposes are displayed.}
\label{fig:setup}       
\end{figure}

The basic setup for the three ECDLs,
described in~\cite{Baillard} is illustrated in
Fig.~\ref{fig:setup}~(a). The emitted light from the laser diode is
collimated by an aspheric lens (CL) ($f$=3.1~mm) with a high
numerical aperture $N.A.$ of 0.68. A part of the collimated light is
back reflected to the laser diode by a partially transmitting mirror
(OC). Different reflectivities $R$ are used for the three lasers,
respectively. Together with the laser diode, the out-coupling mirror
forms the external cavity of the laser with a total length of 70~mm.
The length of the cavity can be modified by displacing the
out-coupler with a piezo-electric transducer (PZT). A higher
stability of the optical feedback is obtained by placing an aspheric
lens (L$_{1}$) with $f$=18.4~mm in front of the out-coupler in focal
length distance (cat eye configuration). The out-coupled light is
then collimated by a second identical lens (L$_{2}$). The
interference filter (IF) is placed inside the cavity. As it was
presented in~\cite{Baillard}, the interference filters have 90\%
transmission and a FWHM of about 0.3~nm. By varying the filter's
angle of incidence relative to the optical axis the wavelength can be
coarsely adjusted. The advantage of this design is that the
wavelength discrimination and the optical feedback are performed by
two independent elements, the interference filter and the
out-coupler~\cite{Baillard}. In combination with the linear design of
the setup, the interference filter-stabilized configuration offers a
higher stability and tunability compared to the Littrow
design~\cite{Zorabedian,Ricci}. Furthermore, the wavelength dependent
spatial displacement of the out-coupled beam is reduced compared to
grating-stabilized set\-ups.

We investigated the implemented laser diodes~\cite{LD-types} for the
three ECDL-systems with respect to their different spectral
properties and their application in the manipulation of atoms. The
laser diode in laser~1 is a common laser diode with a center
wavelength $\lambda_{c}$ of 783~nm and without an AR-coating. With a
reflectivity of 30\% for the out-coupler we obtain stable laser
operation. A laser diode with an AR-coated front facette and with
$\lambda_{c}$=770~nm was chosen for laser~2, providing a broad
wavelength tunability. Here we use a reflectivity of 20\% for the
out-coupler. In laser~3 a DFB-diode with a specified linewidth of
less than 2~MHz and a center wavelength of $\lambda_{c}$=780~nm is
implemented. Due to this intrinsic narrow linewidth we chose an
out-coupler with $R$=20\%.

The design of the tapered laser is illustrated in
Fig.~\ref{fig:setup}~(b). It is based on a self-seeded tapered
amplifier (TA), emitting light in both output directions. The
backward emitted light is focused with an aspheric lens (L$_{1}'$)
with $f$=4.5~mm and $N.A.$=0.55 on a HR coated mirror (M) which
forms together with the front facette of the TA the 77~mm long
external cavity of the TL. We could observe an unstable multi-mode
operation, arising from parasitic feedback of the following
collimation optics, using a common TA-chip with an AR-coating on
both sides. To suppress this disturbing effect we increased the
feedback of the TL-cavity by implementing a TA-chip with a low
reflectivity at the front facette and an AR-coated back
facette~\cite{LD-types}. An interference filter (IF) is set up in
between the mirror and the TA. The back reflected light is seeding
the TA chip. In the gain medium of the TA the light is amplified
and finally emitted from the
front facette. Due to the high asymmetry of the output aperture we
use an aspheric lens (L$_{2}'$) with  $f$=4.5~mm and $N.A.$=0.55 as
well as a cylindric lens (L$_{3}'$) with $f$=80~mm to collimate the
astigmatic output beam. The laser, the beam-shaping optics and a
following optical isolator are shielded by a foamed material to
reduce acoustic noise. For passive stabilization, the laser bodies of
the TL as well as of the other three lasers are milled from a single
solid block of Certal. The block and the diode are independently
temperature controlled.
\section{Characterization of the laser systems}\label{sec:character}
In this section we present and discuss the optical properties of our
laser systems, such as the linewidth and the wavelength tunability.
\begin{figure}[t]
\centering {
  \includegraphics[width=11cm]{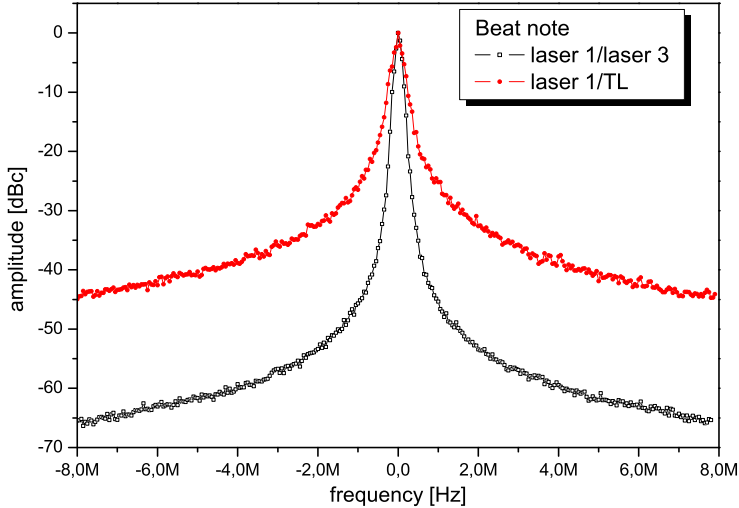}
} \caption{Average of 20 single measurements of the beat signals
of laser~1 and 3 for a sweep time of 2.8~s and a resolution
bandwidth (RBW) of 3~kHz (squares) as well as of the beat between
laser~1 and the TL with a RBW of 30~kHz (dots) for a sweep time of
28~ms (see section~\ref{sec:TLLaser}). The distributions are
normalized with respect to the RBW and are shifted by the
difference frequency.}
\label{fig:beat}       
\end{figure}
We also characterize the output power and the spatial mode
quality. The relevant properties of the ECDL-configurations are
summarized in Table~1. The characterization of the tapered laser
is discussed separately in section~\ref{sec:TLLaser}.
\subsection{Properties of the ECDL-configurations}\label{sec:3Laser}
For the determination of the linewidth of each laser, at 780~nm, a
series of beat measurements have been taken. The beat notes of each
possible combination of two lasers permit an estimation of the
linewidth for each laser. The full width half maximum (FWHM) squared
of the beat signal is given by the sum of the squared linewidths of
two uncoupled lasers. As an example, the beat signal at 410~MHz of
laser~1 and laser~3, both free-running, is shown in
Fig.~\ref{fig:beat}. The instantaneous linewidth can be calculated by
determining the FWHM of a Lorentzian fit of the distribution's flanks
for frequencies with $|\nu-\nu_{0}| > $0.5~MHz. Due to technical
frequency noise the center frequency is distributed randomly and can
be fitted with a Gaussian function for $|\nu-\nu_{0}|<$0.5~MHz.

The values for the linewidths of each free-running laser are
summarized in Table~1. The broadened linewidths for all lasers are
between 120~kHz and 150~kHz determined by the Gaussian fit. We obtain
an instantaneous linewidths for the three lasers of approximately
8~kHz determined via a Lorentzian fit.

\begin{figure}[t]
\centering {
  \includegraphics[width=11cm]{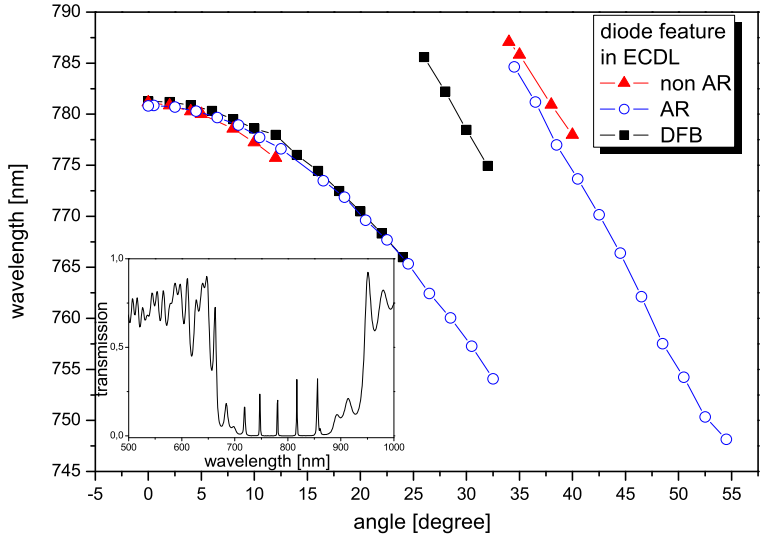}
} \caption{Wavelength tunability of the lasers as a function of
the angle of incidence of the interference filter. The inset shows
the transmission curve of the IF (angle of incidence = $0^\circ$)
measured with a resolution of 1~nm.}
\label{fig:wave_angle}       
\end{figure}

The wavelength of the ECDL can be coarsely adjusted by varying the
angle of incidence of the interference filter relative to the
optical axes or by changing the temperature of the laser diode. A
fine wavelength adjustment can be achieved by changing the
resonator length with the PZT or by modifying the supply current
of the diode. During the measurements we kept the diode's
temperature constant. Changing the filter's angle of incidence
leads to a wide wavelength tunability of several tens of nm as
shown in Fig.~\ref{fig:wave_angle}. The jump of the laser mode
during the wavelength tuning is due to the jump to the adjacent
transmission peak of the interference filter. The inset in
Fig.~\ref{fig:wave_angle} shows the transmission curve of the
interference filter. A broad transmission minimum between 680~nm
and
930~nm is observable. Within that valley, several sharp transmission
peaks separated by about 33~nm arise, limited here by the resolution
(1~nm) of the spectrometer~\cite{Filter}. We achieve wavelength
tunabilities between 11~nm and 32~nm for the three lasers (see
Table~1).

Due to the difference in the physical dimensions of the diodes,
the current dependent frequency tunability differs for the three
configurations, as can be found in Table~1. Stable operation just
on the same diode mode is assured within several tens of GHz for
all lasers. However, mode competition due to the
ECDL-configuration leads of course to mode instabilities which can
be
\begin{table}[t]
\label{tab:properties} \resizebox{\columnwidth}{!} { \centering
\begin{tabular}{|l|c|c|c|}
  \hline
  \ &\textbf{~~laser 1~~}&\textbf{~~laser 2~~}&\textbf{~~laser 3~~}\\
  \hline
  \ LD feature&non-AR  & AR  & DFB\\
  \hline
  \hline
  \ linewidth $\Delta\nu$~[kHz]&130$\pm$25&150$\pm$25 &120$\pm$25\\
  \ (Gaussian)&  &  & \\
  \hline
  \ linewidth $\Delta\nu$~[kHz]&8$\pm$2 &8$\pm$2 &8$\pm$2\\
  \ (Lorentzian)&  &  & \\
  \hline
  \hline
  \ LD current tun. [GHz]& 33 & 43 & 28\\
  \hline
  \ current sen. [MHZ/mA] &92&87&62 \\
  \hline
  \ filter tun. [nm]& 11 & 32 & 19\\
  \hline
  \ PZT tun. [GHz]&2.3&2.3&2.1\\
  \hline
  \hline
  \ $P$~[mW] &34&39&30\\
  \ (@ supply current [mA])&(100)&(110)&(113)\\
  \hline
  \ $I_{threshold}$~[mA] & 23 & 36 & 30 \\
  \hline
  \  estimation TEM$_{00} [\%]$& 99.4&95.8 &99.6 \\
  \ (Gaussian fit)& & & \\
  \hline
\end{tabular}
} \caption[properties]{Optical and spectral properties of the
characterized ECDL-configurations.}
\end{table}
compensated by changing the resonator length. By varying only the PZT
voltage, single-mode operation for the three lasers can be assured
within a span of about 2~GHz, limited by the length of the external
cavity. With the combination of PZT and current tuning, a
mode-hop-free tuning range of several GHz can be accomplished.

A further important property is the output power $P$ of the lasers
which depends on the properties of the implemented diode as well
as on the feedback which is mainly determined by the reflectivity
of the out-coupler. Since the linewidth also depends on $R$, it is
necessary to make a compromise between small linewidth and high
output power. For an optimal operation current we achieve output
powers between 30~mW and 39~mW, measured before going through an
optical isolator. The values for the power as well as the values
for the threshold of the three lasers are summarized in Table~1.

Another criterion of the lasers is the spatial mode profile which is
necessary for a well-defined manipulation of atoms or for the
efficiency of fiber coupling. We characterized the profile by
measuring the beam shape with a CCD-camera at a distance of about
20~cm from the lasers. By fitting a Gaussian function to the
distribution's cross-section, we could coarsely estimate the purity
of the spatial mode profile. All lasers provide a major fraction of
the TEM$_{00}$ mode in their profile. The best performance could be
realized with laser~3, where the distribution corresponds with 99.6\%
to a Gaussian fit. Furthermore, we reach efficiencies of over 72\%
for laser~1 and~3 for the injection of the laser beam into a
single-mode fibre. We inject 63\% for laser~2, where the AR-coating
is responsible for a slightly worse spatial mode shape.

Though the lasers are very similar, they offer different advantages.
Laser 1 is an economic and attractive alternative, due to the low
costs of the implemented laser diode. The large wavelength tunability
of laser~2 makes this system interesting for experiments including
dual species manipulations like atomic potassium (766~nm) and
rubidium (780~nm). Laser~3 offers slightly better properties compared
to the other two lasers concerning the linewidth and spatial mode
profile.

We also measured the linewidth of a laser similar to laser~3 but
without the interference filter. In this case we obtained a linewidth
which is two times bigger than for the IF-stabilized laser. Mode
instabilities due to a possible conflict of the two frequency
selective elements in the design of laser~3, the filter and the
integrated grating structure in the diode, have not been observed.

From the accomplished characterization, we can summarize that all the
three lasers are versatile tools for experiments concerning the
precise manipulation of atoms. Due to their brilliant spectral
properties, the utilization as a Raman-laser~\cite{Cheinet} or for
detection purposes are interesting applications for these lasers.

\subsection{The Tapered Laser}\label{sec:TLLaser}
As we will discuss in the following, the self-seeded tapered
amplifier prototype offers an attractive alternative to other high
power diode laser systems. To deduce the linewidth of the TL we
measured the beat note, shown in Fig.~\ref{fig:beat}, between the
free-running TL and laser~1, locked to a Rb-transition. With the
Gaussian fit of the center region we obtain a FWHM of 200~kHz and
thus a broadened linewidth of $187$~kHz~$\pm12$~kHz for the tapered
laser. Here we assume a broadened linewidth for laser~1 of less than
$100$~kHz, due to the locking of the laser to an atomic transition.

For the instantaneous linewidth of the TL we obtain less than
$85$~kHz from the Lorentzian fit of the distribution's flanks. For
comparison, a commercial self-seeded grating-stabilized TA laser
design for instance offers a linewidth of 1~MHz (specified at 1~ms
measuring time)~\cite{Stry}. Furthermore, we can see in
Fig.~\ref{fig:beat} that for frequencies higher than 5~MHz with
respect to the carrier frequency the background, which is dominated
by the contributions of the amplified spontaneous emission, is 40~dB
smaller compared to the carrier signal.

Due to the cavity length of 77~mm, the mode-hop-free tunability
with the PZT is about 2~GHz. The sensitivity of the laser
frequency for current variations was determined to 19~MHz/mA. A
sweep of several GHz without mode-hopping can be observed for a
combined PZT-current tuning. The specified current noise of the
current driver which is about $\approx5\mu A$ converts with the
measured frequency sensitivity for current variations of 19~MHz/mA
into a frequency noise in the order of several tens of kHz. Thus
the dominant contribution for the broadening of the linewidth is
the noise of the current driver.

The power characteristic is a remarkable feature of the new
TL-design. High output power of 1.24~W at a supply current of
$I_{TA}=2~$A before going through the optical isolator has been
reached. We obtain a slope of 1081~mW/A and a threshold current of
$I_{th}=0.9$~A for the power characteristic.

The spatial mode profile is similar to that of a MOPA-system. This
turns out from the comparable injection efficiencies into a single
mode fibre which is in the case of the TL approximately 51\%. Similar
injection values have been achieved with MOPA-systems~\cite{Voigt}.

For a further consideration we discuss the feedback properties of the
TL. As it was introduced in~\cite{Baillard} the feedback $F$ depends
strongly on the waist $w_{0}$ at the mirror's surface. The feedback,
normalized by the reflectivity, against a disturbance of the mirror
with respect to the optical axes is given by
\begin{equation}
F=e^{-(\alpha\pi w_{0}/\lambda)^{2}}
\end{equation}for a small tilt with the angle $\alpha$ and by
\begin{equation}
F=\left(1+\frac{\delta^{2}\lambda^{2}}{\pi^{2} w_{0}^{4}}\right)^{-1}
\end{equation}
for a displacement $\delta$. According to the formula, the reduction
of the feedback due to a tilt increases with an increasing waist
size, what sets an upper limit for $w_{0}$. On the other hand $w_{0}$
must not be chosen to small to assure stability of the feedback
against displacements of the mirror. A waist of 10~$\mu$m as realized
in the ECDL-configurations as well as a waist of about 1~mm for a
collimated beam in the TL-cavity leads to an unstable, multi-mode
operation of the TL. As an optimal compromise we realized a waist of
about~40~$\mu$m at the mirror's surface. Disturbances of $F$ due to a
tilt of the mirror are negligible for this value of $w_{0}$.

The new laser system offers many advantages compared to
grating-stabilized self-seeded TA-systems or MOPA-systems. In the
tapered laser the interplay between angular and displacement
sensitivity is easier to control compared to Littrow
designs~\cite{Baillard} by choosing an adequate beam waist. A further
advantage of the IF-based system is that the wavelength dependent
spatial displacement of the beam for a tuning of the interference
filter, due to the small length which is approximately~0.5~mm of the
IF, is strongly reduced. It is even more reduced than in the
presented ECDL-configuration, due to the out-coupling at the front
facette of the TA. In addition, the TL offers a more compact and
simplified setup with a base area of 16$\times$7~cm$^{2}$, providing
lower costs compared to conventional MOPA-lasers, where a master and
an amplifier system are required.
\section{Conclusion}
\label{sec:conclusion} In the presented work compact and robust laser
configurations have been realized based on the wavelength
discrimination via narrow-band interference filter. The novel tapered
laser with its high output power of more than 1~W combined with best
spectral properties, reaching a linewidth of less than 85~kHz, offers
a very promising alternative to state-of-the-art systems. The
presented laser systems are currently implemented in atom optical
experiments for cooling and trapping purposes of different atomic
species (Rb,K).
\section{Acknowledgements}
We are grateful to P.~Rosenbusch for detailed information on the
SYRTE ECDL-system and to F.~Scholz from eagleyard Photonics for
technical assistance. The work is supported by the SFB 407 of the
''Deutsche Forschungsgemeinschaft'' and the cooperation FINAQS of the
European Union.


\begin{thebibliography}{}
\bibitem{LD-types} The implemented laser diodes are the following: Sharp: laser~1: GH0781JA2C, eagleyard Photonics: laser~2: EYP-RWE-0790-04000-0750, laser~3: EYP-DFB-0780-00080-1500, TL: modified TA-chip at 780nm;
for the exact value of the reflectivity of the front facette, we
would like to refer to eagleyard Photonics. The interference
filter which has been used in all lasers is a custom-made product
of Research Electro-Optics.
\bibitem{Filter} A spectral analysis of the interference filter has been performed in the Laser Zentum
Hannover by I.~Balasa.
\end{thebibliography}
\end{document}